\documentclass[epj]{svjour}
\usepackage{amssymb}
\usepackage{amsmath}
\usepackage{graphicx}
\usepackage{appendix}
\usepackage{multirow}

\def \var {\textrm{var}}
\def \cov {\textrm{cov}}
\begin{document}
\title{Correlation Testing for Nuclear Density Functional Theory}
\author{M.G. Bertolli
}                     
\offprints{mbertolli@lanl.gov}          
\institute{Theoretical Division, Los Alamos National Laboratory, Los Alamos, NM 87545, USA}
\date{Received: date / Revised version: date}
%
\abstract{
Correlation testing provides a quick method of discriminating amongst potential terms to include in  a nuclear mass formula or functional and is a necessary tool for further nuclear mass models; however a firm mathematical foundation of the method has not been previously set forth.  Here, the necessary justification for correlation testing is developed and more detail of the motivation behind its use is given.  Examples are provided to clarify the method analytically and for computational benchmarking.  We provide a quantitative demonstration of the method's performance and short-comings, highlighting also potential issues a user may encounter.  In concluding we suggest some possible future developments to improve the limitations of the method.
} 
\maketitle

\section{Introduction}

Nuclear structure lies at the heart of many crucial problems 
from nuclear technology to stellar nucleosynthesis, and its theories 
provide a means to explore those systems yet unattainable experimentally.  
A fundamental aspect of nuclear structure, nuclear masses, has had a rich history 
of study.  Since the first work by Bethe and Weizs\"{a}cker \cite{Bethe1936}, nuclear theory 
has sought to provide a consistent and global description of nuclear masses in a concise form.  
To this end, much progress has been made in the more than seventy year history.  
Global approaches include ab intio approaches such as no-core shell model and 
Green function  Monte Carlo \cite{Wiringa2000,Pieper2001}, mean-field  approaches such as 
Skyrme-Hartree-Fock-Bogoliubov and density functional theory~\cite{Stone2007,unedf,unedf2,Goriely2002,Kortelainen2010,Bertolli2012,Bender2003,Stoitsov2003}, microscopic-macroscopic 
approaches such as the finite-range droplet model (FRDM) \cite{Moller1995}, and formulaic 
approaches such as the Duflo-Zuker mass formula \cite{Duflo1995} (for a review see Ref.~\cite{Lunney2003}).  
In particular this work addresses those approaches such as the Duflo-Zuker mass formula 
or density functional theory of Refs.~\cite{Kortelainen2010,Bertolli2012}, as these methods rely 
on the development of a function or functional to accurately describe the relationship between 
nuclear masses and a set of appropriate degrees of freedom.

Indeed, nuclear density functional theory (DFT) has proven to be a useful tool for globally describing 
the ground-state properties of nuclei~\cite{Goriely2002,Bender2003,Lunney2003,Stoitsov2003}.  This
approach is based on the pioneering works of Skyrme~\cite{Skyrme1956} and
Vautherin and Brink~\cite{Vautherin1972,Vautherin1973} and finds theoretical basis on the
theorems by Hohenberg and Kohn~\cite{Hohenberg1964}.  It has been applied
through the self-consistent mean-field computations with
density-dependent energy functionals~\cite{Kohn1965}.  Recently, developments in
global nuclear energy-density functionals have shown great progress.  The Skyrme-Hartree-Fock-Bogoliubov mass
functionals by Goriely {\it et al.}~\cite{Goriely2009} achieve a least-squares
error of about $\chi = 0.58$~MeV to nuclei with $N,Z\geq 8$.  The UNEDF functionals of Ref.~\cite{Kortelainen2010} 
have a least-squares deviation on the order of $\chi=0.97$~MeV and $\chi=1.46$~MeV.  A review on this matter, is 
available in Ref.~\cite{Stone2007}.

While DFT allows for a simple solution to the quantum many-body problem, the construction of the functional itself remains a central challenge in nuclear physics~\cite{unedf,unedf2}. Various efforts aim at
constraining the functional from microscopic interactions~\cite{Stoitsov2010} or devising a systematic approach~\cite{Carlsson2008}.  Many approaches are emperical in nature~\cite{Dobaczewski1984,Oliveira1988,Anguiano2001,Duguet2008}, with 
guidance found from energy-density functionals for dilute Fermi gases with short-ranged interactions~\cite{Furnstahl2007,Puglia2003,Bhattacharyya2005}.  
Refs.~\cite{Bertolli2012,Carlson2003,Papenbrock2006,Bhattacharyya2006,Bulgac2007,Bertolli2008} further suggest simple scaling arguments may guide the form of the functional.  

Bertsch {\it et al.}~\cite{Bertsch2005} employed eigen decomposition to study the importance of 
various linear combinations of terms that enter the functional, drawing guidance from a detailed numerical analysis of the 
functional. This method identifies the relative importance of possible combinations of terms and truncates search directions that are 
flat in the optimization space.  Terms of similar importance redundantly represent a vector in the optimization space.  While an eigen decomposition can be numerically costly, the correlation test discussed in this paper provides similar guidance on term redundancies.  This method avoids analytical difficulties in identifying repeated terms in the optimization space and can be implemented with minimal computational expense.

A future for nuclear mass modeling lies in the further development of functionals.  As the number of terms in functionals increases there exists a real need to generate and select terms systematically, with insight that will aid in minimizing deviations from experiment (see Eq.~(\ref{eq:chi2}) below).  The correlation test was demonstrated in Ref.~\cite{Bertolli2012} in the development of an occupation-number based energy functional for nuclear masses.  In this paper we demonstrate the ability of this approach to provide algebraic insight to the development of nuclear DFT functionals with extremely low computational expense, even while it may not elucidate any direct physical properties.  In Sect.~\ref{proof} we briefly describe the method of Ref.~\cite{Bertolli2012} and provide its motivation along with an analytical example of its performance in identifying algebraic redundancies.  In Sect.~\ref{results} some quantitative results will demonstrate the performance of the method and provide a computational characterization.  Sect.~\ref{limitations} discusses the limitations of correlation testing in developing functionals and in Sect.~\ref{conclusions} conclusions will be made, suggesting some ways in which the method may be improved.   Finally, in the appendix we provide a formal mathematical foundation of correlations as an indication of linear independence.  

\section{Motivation and Background}
\label{proof}

One may view the functional as a solution to the least-squares problem,

\begin{equation}
\min_f (\chi^2) = \min_f \left(\frac{1}{N_{\textrm{pts}}} \sum_{i=1}^{N_{\textrm{pts}}}  \left(f(\mathbf{x}_i)-E^{\textrm{exp}}_i \right)^2 \right)
\label{eq:chi2}
\end{equation}

\noindent where terms of the functional are vectors in the functional space.  In Eq.~(\ref{eq:chi2}) $f(\mathbf{x}_i)$ is the theoretical ground-state energy of the $i^{\textrm{th}}$ nucleus 
provided by the functional, where the functional is fit to $N_{\textrm{pts}}$ experimental ground-state energies, $E^{\textrm{exp}}$.  Insuring the terms of the functional are unique directions in the 
functional space will provide the best solution to Eq.~(\ref{eq:chi2}), for a given number of terms.

The technique of projecting out linearly dependent vectors is not new to the field of optimization.  The principal axis method is frequently employed in the optimization of functions for which derivatives are prohibitively expensive to calculate or unavailable~\cite{Brent1973,Powell1964}, and is based on the concept of orthogonal search directions.

The principal axis method is rooted in Taylor's theorem, whereby any function may be approximated as a quadratic, about some point $\mathbf{x}_0$~\cite{Powell2007},

\begin{eqnarray}
f(\mathbf{x}) &\simeq& f(\mathbf{x}_0)+(\mathbf{x}-\mathbf{x}_0)^T J(\mathbf{x}) \nonumber \\
&+&\frac{1}{2}(\mathbf{x}-\mathbf{x}_0)^T \nabla^2f(\mathbf{x})\bigg|_{\mathbf{x}_0}(\mathbf{x}-\mathbf{x}_0),
\end{eqnarray}

\noindent where $J(\mathbf{x})$ is the Jacobian of $f(\mathbf{x})$, and the principal axis theorem which states any quadratic form can be put in the form $Q(\mathbf{x})$~\cite{Strang1994}

\begin{equation}
Q(\mathbf{x})=\mathbf{x}^TA\mathbf{x},
\label{eq:theorem}
\end{equation}

\noindent where $A$ is an orthogonally diagonalizable symmetric matrix.  

In the principal axis method a set of search directions $\mathbf{u}=u_1,\dots,u_n$ in the $n$-dimensional function space are updated iteratively to provide a new set of search directions $u'_1,\dots,u'_n$ such that after $n$ iterations all $u'_i$ are mutually orthogonal with respect to a quadratic form of the function~\cite{Brent1973,Powell1964}.  

The orthogonal search direction contains no linear dependence and is obtained from an eigen decomposition of $A$~\cite{Strang1994}.   This search direction corresponds to the orthonormal eigenbasis of $A$~\cite{Strang1994} and  minimizes the quadratic form exactly.  For a proof see Ref.~\cite{Powell1964}.  Reference~\cite{Brent1973} employs a singular value decomposition on the search direction $\mathbf{u}$ to insure linear dependence is projected out, citing greater efficiency over eigen decompositions.  The method has been used extensively in many fields including psychology, image processing, and machine learning~\cite{Rusinek1993,Hu2006,Gegenfurtner1992}.  It is available as packaged optimization routines such as {\tt PRincipal AXIS} ({\tt PRAXIS})~\cite{Brent1973}.  In nuclear physics,  the authors of Ref.~\cite{Bertsch2005} have utilized the concept of principal axes when treating the terms of a functional as vectors in the functional's space, again identifying independence through an eigen decomposition.  These well-known approaches, however, can be prohibitively expensive when one must consider searching through hundreds, if not thousands, of possible terms to include in a functional.  They nevertheless served as the initial motivation underlying the method of correlation testing.

The principal axis method provides an effective and fast method for optimization.  Due to the computational expense of minimizing mass models, and the clear direction of model development to include more terms, there is a need for a well-demonstrated way to employ the utility of linear independence in optimization for the selection of new terms.  As will be discussed in the following sections correlation testing provides just such a method.  By using the correlation coefficient as an indication of linear independence amongst a set of terms we are able to select future terms for a functional in a way that provides functional insight and low computation time.  We start by describing in detail the correlation test in Sects.~\ref{proof3} and~\ref{proof2}.

\subsection{Selection of terms based on correlation}
\label{proof3}

We first briefly describe the method detailed in Ref.~\cite{Bertolli2012} for selecting new terms to be included in
the functional.  We will deal only with the aspect of the method utilizing 
the correlation coefficient as a measuring rod for a new term.  While the case in Ref.~\cite{Bertolli2012} is more complex, 
the following nevertheless provides a firm foundation for the use of correlation coefficients in functional analysis.

Consider a functional $F_0(c;\mathbf{x})$, which has $M$ terms $f_1,\dots,f_M$ 
that depend on some set of variables $\mathbf{x}$ and parameters $\{c\}$.

\begin{equation}
 F_0(c;\mathbf{x}) = \sum_{\alpha=1}^{M} c_\alpha f_\alpha(\mathbf{x}) .
  \label{func_noiso}
\end{equation}

Consider the addition of a term $c_{\textrm{new}} f_{\textrm{new}}$, with new fit
parameter $c_{\textrm{new}}$, to the functional. The method is based on the expectation that
the addition of the term $c_{\textrm{new}} f_{\textrm{new}}$ to $F_0$ will be useful in lowering
the chi-square only when it is independent of the $M$ terms already included in the functional.

This independence is defined through the correlation coefficient

\begin{equation}
  R_{f_\alpha,f_{new}}=\frac{\mbox{cov}(f_\alpha,f_{new})}{\sqrt{\var({f_\alpha}) \var({f_{new})}}} . \label{eq:corr}
\end{equation}

Here the covariance is

\begin{equation}
  \cov(f_\alpha,f_{new}) = \langle f_\alpha f_{new} \rangle 
- \langle f_\alpha\rangle \langle f_{new} \rangle ,  
\end{equation}

and the average $\langle \cdot \rangle$ is computed with respect to the 
$N_{\rm pts}$ experimental points available for Eq.~(\ref{eq:chi2}).  The variances, $\var (f_i)$, are then

\begin{eqnarray}
  \var({f_{new}})&=& \cov (f_{new},f_{new}) \nonumber \\
                           &=&\langle f_{new}^2\rangle -\langle f_{new}\rangle^2
 \end{eqnarray}
 
 \noindent and
 
\begin{eqnarray} 
  \var({f_\alpha})&=& \cov (f_\alpha,f_\alpha) \nonumber \\
                            &=&\langle {f_\alpha}^2\rangle -\langle {f_\alpha}\rangle^2. 
\end{eqnarray}

We note that the correlation coefficient is independent of the coefficient $c_{\textrm{new}}$ of the new term under consideration.  Should the correlation be sufficiently low for all included terms, the new term $c_{\textrm{new}} f_{\textrm{new}}$ may be included in the functional.  This allows one to investigate many different forms of functional terms  without the time-consuming and computationally expensive aspects of performing a full minimization of the least-squares for each new term under question.  It is an application of the already known correlation analyses to a function but {\it prior} to fitting.

\subsection{Correlation as an indication of independence}
\label{proof2}

The concept of linear independence is familiar to the realm of functional optimization.  Where the functional is a solution to Eq.~(\ref{eq:chi2}) the concepts can be seen as meaningful in the development of mass models~\cite{Bertolli2012,Bertsch2005}.  Yet to justify the use of the method in Ref.~\cite{Bertolli2012} one must ask, is the correlation coefficient Eq.~(\ref{eq:corr}) a meaningful indication of the linear independence between two functions?  This question becomes particularly important when one recalls the frequent warning to new students of statistics that a zero covariance does {\it not} imply independence~\cite{Smith1991}.  While this statement may seem to condemn the method outright, we will address the subtlety that while a zero covariance does not imply statistical independence it does in fact indicate linear independence of functions of multiple variables in the linear algebraic sense.  Here, we use a simple yet demonstrative case to show the practical output of the correlation test.  In the appendix, we set out the necessary formal foundation through a proof by contradiction.  In Sect.~\ref{results} we provide detailed benchmarks and further define the specifics of the correlation test in nuclear mass model development

Where terms are highly correlated the variation of the function with respect to fit coefficient may be absorbed by the simultaneous variation of the other term coefficients.  As an example we consider a two-term function of total proton ($Z$) and neutron ($N$) number of a nucleus and fit coefficients $c_1,c_2$

\begin{equation}
g(N,Z) = c_1(N^2-Z^2)+c_2NZ,
\label{eq:testfunc}
\end{equation}

\noindent where one may interpret the terms as incorporating neutron-neutron, proton-proton and neutron-proton two-body interactions, respectfully.  Each such interaction is physically motivated and so one may conclude that both terms of $g(N,Z)$ should be included in the description of nuclei with different leading coefficients $c$.

We can construct the correlation matrix between the two terms in $g(N,Z)$ via Eq.~(\ref{eq:corr}), where the statistical averages are over a set of about 2000 nuclei ranging across the entire nuclear chart.  The resulting $2\times2$ correlation matrix is given in Table~\ref{table:testfunc_corr}.

\begin{table}[!ht]
\begin{center}
\caption{Correlation matrix for the function $g(N,Z)$.  The high correlation indicates a redundancy in the function space.\label{table:testfunc_corr}}
\begin{tabular}{|c|c c|}
  \hline
  $N^2-Z^2$&1 & 0.99 \\
  $NZ$&0.99& 1 \\
  \hline
\end{tabular}
\end{center}
\end{table}

\noindent The correlation matrix in Table~\ref{table:testfunc_corr} indicates that the two terms are redundant, which may not have been clear from initial physical considerations.  As such, in fitting we may reduce the number of coefficients.

\begin{equation}
g(N,Z) = c_t(N^2-Z^2+NZ)
\label{eq:testfunc2}
\end{equation}

\noindent where $c_t$ now absorbs the simultaneous variation of the $c_1,c_2$.  Indeed for this case we can show the validity of Eq.~(\ref{eq:testfunc2}) analytically.  We perform a coordinate rotation from $N,Z \rightarrow N',Z'$ by $\theta = \pi/4$:

\begin{eqnarray}
 \begin{pmatrix}
  N' \\
  Z'
 \end{pmatrix} &=&
 \begin{pmatrix}
  \cos \theta & -\sin \theta \\
  \sin \theta & \cos \theta \\
 \end{pmatrix}
 \begin{pmatrix}
  N \\
  Z
 \end{pmatrix} \nonumber \\
 &=& \begin{pmatrix}
   N\cos \theta -Z\sin \theta \\
  N\sin \theta+Z\cos \theta
 \end{pmatrix}. 
\end{eqnarray}

\noindent Where the function surface is invariant under rotations about the the dependent axis $g(N,Z)=g'(N,Z)$, we can write the first term $N^2-Z^2$ equivalently in terms of the new coordinates $N',Z'$

\begin{equation}
(N\cos \theta -Z\sin \theta)^2-(N\sin \theta+Z\cos \theta)^2 = -2NZ
\label{eq:demonstrate}
\end{equation}

\noindent for $\theta = \pi/4$.  So we find that through a simple rotation the first term $N^2-Z^2$ is exactly equivalent to the second term $NZ$ to within a constant.  Therefore we can equivalently cast Eq.~(\ref{eq:testfunc}) as

\begin{eqnarray}
g(N,Z) &=& c_1(N^2-Z^2)+c_2NZ \nonumber \\
&=& NZ(-2c_1+c_2),
\label{eq:finalexample}
\end{eqnarray}

\noindent where the coefficient $-2c_1+c_2$ can be replaced with $c_t$ as in Eq.~(\ref{eq:testfunc2}).  In the limit of an infinite number of nuclei, the statistical correlation between the two terms will be 1.0, as expected when the functions differ by only a constant.  This simple example has provided an clear demonstration of the underlying concept in correlation testing, where the redundancy indicated by the correlation matrix is confirmed analytically.  Correlation testing is valuable where simple algebraic manipulation is not possible to uncover repeated terms.  We now turn to a more rigorous characterization of the performance of the correlation test across many terms.

\section{Performance of Correlation Test}
\label{results}

While we have provided motivation for the correlation test, along with the necessary mathematical foundation for the general case, it is often helpful to view simple examples for quantitative insight.  In this section we will address the potential benefits of using correlation testing in functional development.  The greatest appeal of the correlation test lies in its inexpensive computational costs.  As such, we benchmark the performance of the method by looking at the cost of a full minimization of a set of terms versus selecting terms from the correlation test.   As was done in the development of the full energy functional in Ref.~\cite{Bertolli2012} we perform correlation tests on an initial set of about 200 different terms and utilize the same definition for high and low correlation.  In the context of nuclear mass models we define ``high" correlation as where the absolute value of the correlation coefficient $R$ between the new term and the pre-existing terms is  greater than 0.5, $|R|>0.5$.  Likewise, ``low" correlation is where the absolute value of the correlation coefficient between the new term and any pre-existing term is never greater than 0.5, $|R|<0.5$.  Throughout the paper we will make use of the average absolute value of the correlation coefficient, $\langle |R| \rangle$, for brevity.  We obtain $\langle |R| \rangle$ by averaging over the absolute value of the correlation coefficient between the new term and all terms already present:

\begin{equation}
\frac{1}{M}\sum_\alpha^M |R_{f_\alpha,f_{new}}|. 
\label{eq:avR}
\end{equation}

\noindent While Eq.~(\ref{eq:avR}) gives an average, we emphasize that when describing ``high'' or ``low'' correlation all individual correlation coefficients between the new term and any term already present are also always either within the regime of ``high" or ``low".

As the appropriate interpretation of the correlation coefficient is not readily generalized across all systems, we employ a training step similar to those used in machine learning for the particular system in question~\cite{Mitchell1997}.  The appropriate interpretation of ``high" and ``low" correlation is determined empirically from performance of the test over a set of training terms.  The performance of the mass model is measured by how well it minimizes the $\chi^2$ value of Eq.~(\ref{eq:chi2}).  Therefore, we measure the performance of an individual term $f_{new}$ by the amount $\Delta \chi$ the solution to Eq.~(\ref{eq:chi2}) is reduced when added to the functional in question.  

\begin{equation}
\Delta \chi = \chi(f(\mathbf{x}))-\chi(f(\mathbf{x})+f_{new}(\mathbf{x}))
\end{equation}

\noindent Here, a higher $\Delta \chi$ will indicate a better performance, as the addition of the term $f_{new}$ results in a greater reduction of $\chi$ and thus better solution to Eq.~(\ref{eq:chi2}).  

\begin{figure}
  \includegraphics[width=0.5\textwidth,angle=0]{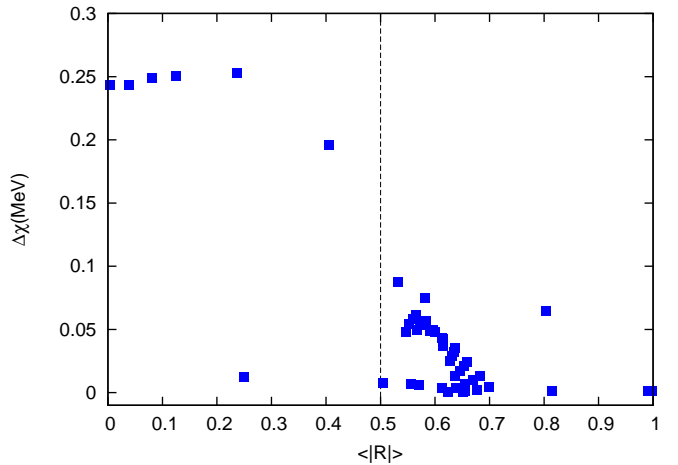}
  \caption{(Color online) Plot of the change in least-squares error, $\Delta \chi$, resulting from 
  the addition of a new training term as a function of the term's average correlation coefficient 
  $\langle |R| \rangle$.  The training terms are used to determine the appropriate cut-off between 
  ``high'' and ``low'' correlation.  The dotted black line indicates the cut-off 
  between high and low correlation.}
 \label{fig:deltachi_training}
\end{figure}

Figure~\ref{fig:deltachi_training} demonstrates the performance of a term as a function of correlation.  Here the training set of terms is taken to be roughly 20\% of the full set of terms to be tested for inclusion in the model~\cite{Mitchell1997}.  A clear distinction exists between the performance of terms with $|R|<0.5$ and  $|R|>0.5$, indicating this as the appropriate cut-off between ``high" and ``low" correlation in this system.  The presence of such a cut-off follows from the principal axis method presented in Sect.~\ref{proof}, where the final orthogonal search direction defines vectors with no linear dependence.  A proof of the correlation coefficient as an indication of linear independence is given in the Appendix.  A detailed discussion of the difficulty of a general interpretation of correlation is provided in Sect.~\ref{doi}.

For our demonstration we take a simplified version of the energy functional in Ref.~\cite{Bertolli2012}, with fit coefficients $\mathbf{c}=\{c_c,c_s,c_{as},c_{ss},c_1,c_2,c_3\}$:

\begin{equation}
\begin{split}
F (c; n, z)&=  c_c \frac{Z(Z-1)}{A^{1/3}}+\hbar \omega \bigg( c_1A^{-1/3}\sum_k k (z_k+n_k)  \\
&+c_2 A^{-1/3}\sum_k \left( \frac{z_k(z_k-1)}{k} +\frac{n_k(n_k-1)}{k} +\frac{2 n_k z_p}{k} \right) \\
& +c_3 A \bigg),
\end{split}
\label{eq:func0}
\end{equation}

\noindent where $z_k$ and $n_k$ are the occupations of proton and neutron shells, respectively, $A=N+Z$ is the atomic number of the nucleus, and

\begin{eqnarray}
\hbar \omega(c_s,c_{as},c_{ss}) &=& 1-c_s A^{-1/3}\nonumber \\
  &&-\frac{c_{as}}{1+c_{ss}A^{-1/3}} \frac{T(T+1)}{A^2}. \label{hbarcomplex}
\end{eqnarray}

\noindent In Eq.~(\ref{hbarcomplex}), $T=|N-Z|/2$ defines the isospin.  This energy functional employs a spherical harmonic oscillator basis, whose eigenvalue solutions are known to be dependent on the oscillator frequency $\omega$ and the oscillator shell $k$.  The principal quantum number $k$ is related also to the dimensionality of the shell $d_k = (k+1)(k+2)$ and thus is used in the functional to ensure appropriate scaling of terms.  For a detailed discussion on $\hbar \omega$ and scaling we refer the reader to Refs.~\cite{Bertolli2012,Bertolli2011}.  

Our truncated form of the energy functional is taken so as to include one- and two-body terms along with the Coulomb interaction and a volume term $c_3A$.  Furthermore, as we are interested in demonstrating the use of correlation testing, we treat $F(c;n,z)$ as a mass formula and set the occupations $n,z$ at a na\"{i}ve level filling.  This removes the multilevel optimization necessary to fit the full functional~\cite{Bertolli2012}.  While for the remainder of this section we will proceed with a formula, the demonstration is extendable to functionals.

In the optimization of Eq.~(\ref{eq:chi2}) the addition of a new term and fit coefficient $c_\textrm{new} f_\textrm{new}$ can never result in raising the least squares error.  That is

\begin{equation}
\begin{split}
\min_c &\left( \frac{1}{N_{\textrm{pts}}} \sum_{i=1}^{N_{\textrm{pts}}}  \left( F(c;n_i,z_i)-E^\textrm{exp}_i \right)^2 \right) \geq  \\
&\min_c \left(  \frac{1}{N_{\textrm{pts}}} \sum_{i=1}^{N_{\textrm{pts}}}  \left( F(c;n_i,z_i)+c_\textrm{new} f_\textrm{new}-E^\textrm{exp}_i \right)^2 \right).
\end{split}
\label{eq:min}
\end{equation}

\begin{figure}
  \includegraphics[width=0.5\textwidth,angle=0]{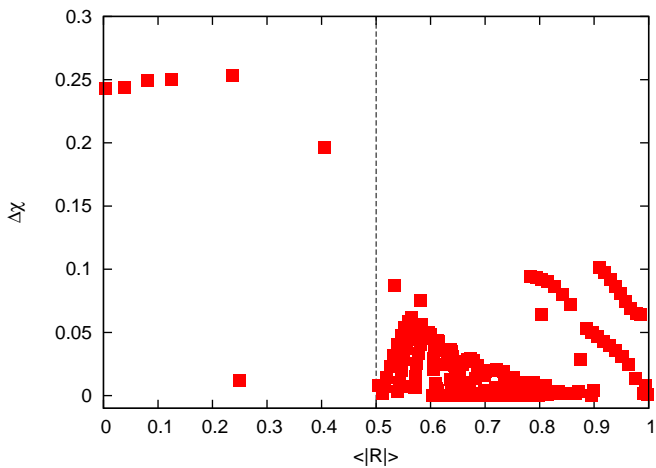}
  \caption{(Color online) Plot of the change in least-squares error, $\Delta \chi$, resulting from 
  the addition of a new term as a function of the term's average correlation coefficient 
  $\langle |R| \rangle$.  A large number of highly correlated ($\langle |R| \rangle>0.5$) terms 
  result in no change to the least-squares error.  The dotted black line indicates the cut-off 
  between high and low correlation.  The subset of terms with high correlation which contribute 
  to a $\Delta \chi \simeq 0.1$ MeV is discussed in Sect.~\ref{nophysics}.}
 \label{fig:deltachi}
\end{figure}

\noindent Equation~(\ref{eq:min}) imposes the inequality because the optimization will result in equality, with $c_\textrm{new}=0$,  when the addition of a term does not create a lower minimum.  We perform a full minimization over each new functional 

\begin{equation}
F (c; n, z)+c_if_i
\end{equation}

\noindent where $f_i$ is the $i^\textrm{th}$ new term of the set.  In Fig.~\ref{fig:deltachi} we plot the change in least-squares error , $\Delta \chi$, in units of MeV as a function of $\langle |R| \rangle$, Eq.~(\ref{eq:avR}), for all terms reviewed for this paper.  The fit is performed over the set of 2,049 nuclei from the 2003 atomic data evaluation whose uncertainty in the binding energy is below 200 keV ~\cite{Audi2003}.  A summary of the number of terms resulting in a given performance is given in Tables~\ref{table:breakdown} and~\ref{table:breakdown2}.

\begin{table}[!ht]
\begin{center}
\caption{Break-down of how many terms in the full set with ``high'' ($>0.5$) correlation perform at a given level. The left-hand column identifies
 the percentage of the full set of terms (percentages are rounded to nearest integer).  The right-most column gives range of $\Delta \chi$ performance resulting from the inclusion of those terms.\label{table:breakdown}}
\begin{tabular}{ |c|c|}
\hline
\multicolumn{2}{ |c| }{``high'' $\langle |R|\rangle$} \\
\hline
\% of full set & $\Delta \chi$ (keV)\\ \hline
85 & $<50$ \\
39 & $<10$ \\
20 & $0$  \\
\hline
\end{tabular}
\end{center}
\end{table}

\begin{table}[!ht]
\begin{center}
\caption{Break-down of how many terms in the full set with ``low'' ($<0.5$) correlation perform at a given level. The left-hand column identifies
 the percentage of the full set of terms.  The middle column provides a further break-down on how many of those terms with ``low'' correlation 
 give a certain performance (percentages are rounded to nearest integer).  The right-most column gives range of $\Delta \chi$ performance resulting from the inclusion of those terms.\label{table:breakdown2}}
\begin{tabular}{ |c|c|c|}
\hline
\multicolumn{3}{ |c| }{``low'' $\langle |R|\rangle$}  \\
\hline
\% of full set & \% of ``low'' set & $\Delta \chi$ (keV) \\ \hline
\multirow{2}{*}{15} & 86 &  $>150$ \\
 & 14 &  $<150$ \\
\hline
\end{tabular}
\end{center}
\end{table}

Those with a low average correlation coefficient, $\langle |R| \rangle<0.5$, caused the greatest change in $\chi$, significantly higher over changes from terms with high correlation, $\langle |R| \rangle>0.5$.  Of those with low correlation, $\sim$14\% contributed a change in $\chi$ of less than 0.15 MeV.  It is no surprise the majority, 85\%, of all the terms had high correlation and reduced the least-squares error by 50 keV or less, whereas 39\% of the terms reduced the least-squares error by 10 keV or less.  Visible in Fig.~\ref{fig:deltachi}, are a significant number of terms, 20\%, which resulted in a $\Delta \chi = 0$ MeV.  For these terms the coefficient of the added term was fit to $c_i = 0$ and give the equality case of Eq.~(\ref{eq:min}).

In comparing the computational time required to minimize each term individually to selecting terms via a correlation test we find significant advantage in the correlation test.  Minimizing the function for a single term required on the order of 170 CPU seconds, or about one day wall-clock time on a single 3.4 Intel Core i7 processor for the full set of about 200 terms.  In contrast, a correlation test required only 1.62 CPU seconds to return the correlation analysis of all the terms in question.  Where it is crucial to minimize the number of parameters in a model in order to avoid obscuring physical insight~\cite{Danielewicz2009}, the correlation test shows a clear advantage in providing a substantially less time-consuming approach to eliminating large sets of terms under consideration.

\subsection{Detailed Examples}
\label{results2}

While the significantly decreased computational cost over full minimizations for each test functional is a profound advantage of the correlation test, we provide further quantitative results for the reduction in least-squares error when the method is used by considering a simple test case.  From the principal axis method and Refs.~\cite{Bertsch2005,Bertolli2012}, we expect linearly independent terms to reduce the least-squares error more significantly over linearly dependent terms.  We therefore add new terms to Eq.~(\ref{eq:func0}), chosen by testing their correlation to terms already present, and compare the effect of the new terms in lowering the least-squares error as a function of correlation.  From the 200 terms considered above, we have selected four (two with low correlations, two with high correlations) to demonstrate the method's utility.  The two terms chosen with high correlation with terms in Eq.~(\ref{eq:func0}) are:

\begin{eqnarray}
f_1 &=& c_{f1} \sum_k \frac{n_kz_k^2+z_kn_k^2}{k^4}, \label{eq:f1} \\
f_2 &=& c_{f2}\sum_k \frac{(n_k+z_k)|n_k-z_k|}{k^2}, \label{eq:f2}
\end{eqnarray}

\noindent where both terms scale as the total number of nucleons $A=N+Z$.  The new terms chosen with low correlation are:

\begin{eqnarray}
f_3 &=& c_{f3}\sum_k\left(  \frac{\sqrt{d_k}}{2}-\frac{2}{d_k^{3/2}}(z_k-d_k/2)^2 \right),\nonumber \\
  && \times  \sum_\ell \left(\frac{\sqrt{d_\ell}}{2}-\frac{2}{d_\ell^{3/2}}(n_\ell-d_\ell/2)^2 \right),  \label{eq:f3} \\
f_4&=& c_{f4} \frac{\delta}{\sqrt{A}} \label{eq:f4}
\end{eqnarray}

\noindent where $d_k$ is the dimensionality of the $k^{\textrm{th}}$ shell and $n_f, z_f$ are the occupations of the highest occupied neutron and proton shells, respectively.  Here, $\delta$ is 1, 0, and -1 for even-even, odd mass, and odd- odd nuclei, respectively.  Scaling of terms is at most $A$~\cite{Bertolli2012}.  

We now briefly describe the terms in Eqs.~(\ref{eq:f1})-(\ref{eq:f4}) as they relate to a mass formula.  Equation~(\ref{eq:f1}) may be physically motivated for inclusion as a three-body effect, whereas Eq.~(\ref{eq:f2}) may also be physically motivated as a density-dependent symmetry energy~\cite{Vretenar2003,Lagaris1981,Wiringa1988,Dong2012,Shetty2007}.  Likewise, Eq.~(\ref{eq:f3}) may be included in a mass formula to account for two-, three-, and four-body deformation effects as it has a linear onset for almost-empty and almost-filled shells and assumes its maximum at mid-shell with half-filled occupations~\cite{Bertolli2012}.  Equation~(\ref{eq:f4}) is the familiar bulk pairing interaction of the Bethe-Weizs\"{a}cker mass formula~\cite{Bethe1936}.  As such, all terms may be reasonably considered for inclusion in a mass formula, and indeed Eqs.~(\ref{eq:f3}) and~(\ref{eq:f4}) are in the functional of Ref.~\cite{Bertolli2012}.  However, the focus of this paper is the effects of using correlation testing and so we wish to demonstrate the usefulness of  including a term when physical arguments are not considered.  This limitation of the method is discussed in Sect.~\ref{limitations}.

We determine the parameters $\mathbf{c}$ through a least-squares fit of the truncated formula in Eq.~(\ref{eq:func0}) to a set of 2,049 nuclei from the 2003 atomic data evaluation~\cite{Audi2003} whose uncertainty in the binding energy is below 200 keV.  The resulting least-squares error is $\chi  = 1.687$ MeV, with a total of seven fit coefficients.  We add an eighth fit coefficient with the inclusion of one of the above described terms, Eqs.~(\ref{eq:f1})-(\ref{eq:f4}).  The result of their inclusion is summarized in Table~\ref{table:results}, where we show the $\chi$ value obtained from the addition of each individual term to Eq.~(\ref{eq:func0}) and the average absolute value of the correlation coefficient for that term, as calculated by Eq.~(\ref{eq:avR}).  Based on the method of correlation testing (and thus the value of $\langle |R| \rangle$, as shown in Table~\ref{table:results}) we would not expect the addition of the $f_1$ or $f_2$ to result in a significant decrease in the $\chi$ value.  Indeed, adding $f_1$ and refitting to the parameters $c$ to the 2,049 nuclei gives $\chi =  1.686$ MeV, or a reduction of only $\Delta \chi = 0.001$ MeV in the least-squares error.  Adding $f_2$ and refitting the formula gives $\chi = 1.686$ MeV, the same small reduction as with $f_1$.  Thus, based purely on the correlation test the terms $f_1$ and $f_2$ would not be included in the development of a mass formula.

We must now, however, demonstrate that those terms with low correlation {\it are} beneficial to the development of a mass formula.  Proceeding as before, we add $f_3$ and refit the formula obtaining a least-squares error of $\chi = 1.491$ MeV.  This constitutes a reduction in $\chi$ compared to the formula in Eq.~(\ref{eq:func0}) of $\Delta \chi = 0.196$ MeV.  While the formula is still far from competitive,  the reduction is large by mass formula standards~\cite{Lunney2003}.  The addition of $f_4$ gives a similarly dramatic change.  Refitting the formula with $f_4$ gives $\chi = 1.444$ MeV, a reduction in $\chi$ of $\Delta \chi =0.243$ MeV.  Both terms $f_3$ and $f_4$ cause reductions in $\chi$ two orders of magnitude greater than those changes by the highly correlated terms $f_1$ and $f_2$.  Furthermore, the correlation test in principle eliminated the two functions which provided the least reduction, decreasing the set of terms to undergo a full minimization in a method that is less computationally expensive than a singular value or eigen decomposition.  In the case of this formula, the fitting procedure is not costly.  However, limiting the number of candidate terms becomes increasingly important for computationally expensive fits (for a discussion on computational expense see Ref.~\cite{Ng2011}).  The above example provides a demonstration of how the correlation test provides insight into what candidate terms may reduce the least-squares error of a mass formula by the greatest amount without the computational expense of a full minimization.

\begin{table}[!ht]
\begin{center}
\caption{Least-squares deviations of binding energies 
resulting from a global fit of the formula to 2,049 
nuclei~\cite{Audi2003}. The left-hand column identifies the form 
of the formula in terms of which of the four terms $f_1,f_2,f_3,f_4$ 
are added to Eq.~(\ref{eq:func0}).  The right-most column gives the 
average correlation $\langle |R| \rangle$ of the added term relative to terms in Eq.~(\ref{eq:func0}).  
Correlations may be thought of as either low ($<0.5$) or high ($>0.5$).\label{table:results}}
\begin{tabular}{|c|c|c|}
  \hline
  Form & $\chi$ (MeV)&$\langle |R|\rangle$ \\
  \hline
  $F (c; n, z)$&1.687 & -- \\		
  $F (c; n, z)+f_1$& 1.686 &$ \simeq 0.998$ \\
  $F (c; n, z)+f_2$& 1.686 & $ \simeq 0.815$ \\
  $F (c; n, z)+f_3$& 1.491 & $\simeq	 0.405$ \\
  $F (c; n, z)+f_4$& 1.444 & $\simeq 0.004$ \\
  \hline
\end{tabular}
\end{center}
\end{table}

\section{Limitations}
\label{limitations}
On equal footing with the benefits of the correlation test we will also discuss the limitations of the method, which have been indicated in the previous section.  As in Sect.~\ref{results} we will proceed quantitatively in highlighting the possible limitations of correlation testing.

\subsection{Degree of Independence}
\label{doi}

Bertsch {\it et al.}~\cite{Bertsch2005} order the importance of principal axes of the functional by the eigenvalues of their eigen decomposition.  In the case of correlation testing one may be tempted to interpret the absolute value of the correlation coefficient $|R|$ as an indication of the level of dependence between two terms.  However, the authors of Refs.~\cite{Cohen1988,Fulekar2009} warn against such a general interpretation.  There is a need for context when giving meaning to the value of $|R|$ and definitions of ``high" and ``low" correlation are somewhat arbitrary.  References~\cite{Towsley2011} and~\cite{Ross2010} demonstrate the covariance as isomorphic to the cosine of a ``correlation angle" between two statistical quantities when $\var (x_1), \var (x_2) \neq 0$, and $x_1,x_2 \in \mathbb{R}$.  However, the authors again caution that such an interpretation provides no information on the linear dependence we seek to find in functional development.

Figure~\ref{fig:deltachi} similarly indicates this limitation of the correlation test for the full set of about 200 terms.  One may define a vertical division (black dotted line of Fig.~\ref{fig:deltachi}) between ``high" and ``low" correlation within the context of the problem, but the trend below $\langle |R| \rangle=0.5$ is primarily a flat line, not indicating a reliable statement on the degree of independence.  However, the correlation test may be used in a complimentary fashion to the method of Ref.~\cite{Bertsch2005} by significantly reduced the number of terms, making an eigen decomposition numerically feasible.

\subsection{Limited Insight from the Correlation Test }
\label{nophysics}

There are a few other technical issues of which the would-be user must be wary.  The first of these we mention is the inherent linearity of the correlation coefficient.  
The correlation coefficient provides a constant ratio between its arguments and thus is only a measure of the linear relationship between two variables~\cite{Cox1974,Montgomery2003}.  To show this, we consider a different representation of Eq.~(\ref{eq:corr}):

\begin{equation}
R_{f_\alpha,f_{new}} = b_{f_\alpha,f_{new}} \frac{\var (f_\alpha)}{\var (f_{new})},
\label{eq:corr2}
\end{equation}

\noindent where $b_{f_\alpha,f_{new}}$ is the slope of a regression line relating the quantities $f_\alpha$ and $f_{new}$.  Furthermore in Eq.~(\ref{eq:corr}), the act of averaging and scaling (i.e. the factor $1/\sqrt{\var(f_\alpha)\var(f_{new})}$) allows the coefficient to be independent of the scales of the arguments.  The correlation coefficient is then simply a standardized ratio describing a linear (constant) relationship between the two arguments (for details on the linearity, and a derivation of Eq.~(\ref{eq:corr2}), please see Ref.~\cite{Rodgers1988}).

As a result of its linearity, the correlation coefficient contains no knowledge of the non-linear behavior that exists between two variables.  In fact, this linearity contributes to a far more fundamental (and dangerous) problem with the correlation test: it lacks any knowledge of the physics behind the functional or formula being built.  This, perhaps most obvious, issue is manifest in two ways.  We start by quantitatively demonstrating the first: physically important terms may be omitted by the correlation test due to high correlations.

To demonstrate that some terms which may contain important physical properties might be rejected by the correlation test we look to the Bethe-Weizs\"{a}cker mass formula~\cite{Bethe1936} with no pairing, as beyond its historical importance it provides the ground-state energy through four terms of transparent physical meaning:

\begin{equation}
E=a_vA-a_sA^{2/3}-a_c\frac{Z(Z-1)}{A^{1/3}}-a_A\frac{(N-Z)^2}{A}.
\label{eq:Lunney2003}
\end{equation}

\noindent The radius of a nucleus is proportional to $A^{1/3}$, therefore the first term $a_vA$ represents a volume and accounts for the saturation of the nuclear force.  The remaining terms then represent a surface term ($a_s$), a Coulomb contribution ($a_c$), and an isospin term to account for asymmetry between protons and neutrons ($a_A$).  The Bethe-Weizs\"{a}cker mass formula in this form achieves a least-squares error of $\chi \simeq 3.0$ MeV when fit to the set of 2,049 nuclei from the 2003 atomic data evaluation~\cite{Audi2003}.  By looking at correlations amongst the terms in Eq.~(\ref{eq:Lunney2003}) we may reasonably conclude the bulk volume and surface components of the nucleus are fully represented in this function space, and so Eq.~(\ref{eq:Lunney2003}) serves as a suitable reference by which to compare possible new terms.  Table~\ref{table:BWcorr1} shows the correlation matrix for Eq.~(\ref{eq:Lunney2003}), where one can see that all the terms may considered highly correlated with each other. 

\begin{table}[!ht]
\begin{center}
\caption{Correlation matrix for terms of the Bethe-Weizs\"{a}cker mass formula~\cite{Bethe1936} with no pairing.\label{table:BWcorr1}}
\begin{tabular}{|c|c c c c c|}
  \hline
  $A$&1 &&&&\\
  $A^{2/3}$&0.995& 1 &&&\\
  $\frac{Z(Z-1)}{A^{1/3}}$&0.982& 0.968 & 1&& \\
  $\frac{(N-Z)^2}{A}$&0.800& 0.777 & 0.71942 & 1& \\
  \hline
\end{tabular}
\end{center}
\end{table}

Physical arguments motivate the inclusion of these four terms beyond the indicated redundancy of the correlation test.  However, if one would seek to include terms which are not correlated to the four in Eq.~(\ref{eq:Lunney2003}) they may arrive at he fifth term of the full Bethe-Weizs\"{a}cker mass formula, the pairing term:

\begin{equation}
a_P\frac{\delta}{\sqrt{A}},
\label{eq:pair}
\end{equation}

\noindent where $\delta$ is 1, 0, and -1 for even-even, odd mass, and odd-odd nuclei, respectively.  As one can see in Table~\ref{table:BWcorr2}, the pairing term has a low correlation to all other terms in the mass formula, and so is acceptable not only through it's well-known physical motivation, but  by the correlation test.  Its addition to the Bethe-Weizs\"{a}cker mass formula reduces the least-squares error by about 0.1 MeV to $\chi \simeq 2.9$ MeV.  While a greater change in least-squares error may have been expected from a term with such a low correlation, the reduction is meaningful~\cite{Lunney2003} and reasons behind it not being greater are discussed below.

\begin{table}[!ht]
\begin{center}
\caption{Correlation matrix for terms of the full Bethe-Weizs\"{a}cker mass formula~\cite{Bethe1936} with pairing. The pairing term passes the correlation test for inclusion.\label{table:BWcorr2}}
\begin{tabular}{|c|c c c c c|}
  \hline
  $A$&1 &&&&\\
  $A^{2/3}$&0.995& 1 &&&\\
  $\frac{Z(Z-1)}{A^{1/3}}$&0.982& 0.968 & 1&& \\
  $\frac{(N-Z)^2}{A}$&0.800& 0.777 & 0.71942 & 1& \\
 $\frac{\delta}{\sqrt{A}}$&-7$\times 10^{-3}$& -1$\times 10^{-2}$ & 4$\times 10^{-2}$ & -1$\times 10^{-2}$ &1 \\
  \hline
\end{tabular}
\end{center}
\end{table}

\begin{table}[!ht]
\begin{center}
\caption{Average correlation $\langle |R| \rangle$ of the terms in the Bethe-Weizs\"{a}cker mass formula~\cite{Bethe1936}, Eq.~(\ref{eq:func0}), and the $\chi$ value resulting from a fit of the mass formula to the 2,049 nuclei~\cite{Audi2003} with the term removed. \label{table:Lunney2003corr}}
\begin{tabular}{|c|c|c|}
  \hline
  Term & $\langle |R| \rangle$ & $\chi$ (MeV) without term \\
  \hline
  $A$ & 0.7 & 38.04\\
  $A^{2/3}$ & 0.7 & 14.01\\
  $\frac{Z(Z-1)}{A^{1/3}}$ & 0.7& 25.2\\
  $\frac{(N-Z)^2}{A}$ & 0.6 & 23.19\\
  $\frac{\delta}{\sqrt{A}}$ & 0.008 & 3.0\\
  \hline
\end{tabular}
\end{center}
\end{table}

\noindent All of the terms in the  Bethe-Weizs\"{a}cker mass formula are correlated to each other with an average correlation coefficient of at least $\langle |R| \rangle \gtrapprox 0.6$, except for the pairing term for which  $\langle |R| \rangle \simeq 8 \times 10^{-3}$.  Despite the volume, surface, Coulomb and isospin terms having ``high" correlation by the definition used for Eq.~(\ref{eq:func0}) and Ref.~\cite{Bertolli2012} they are of clear physical importance in describing the bulk behavior of nuclei~\cite{Lunney2003}.  Furthermore, their removal increases the least-squares error of the Bethe-Weizs\"{a}cker mass formula from $\chi = 2.9$ MeV to between $\chi = 14-38$ MeV when fitted to the set of 2,049 nuclei~\cite{Audi2003}.  Details of the effect of each term on $\chi$ is give in Table~\ref{table:Lunney2003corr}.  The lack of impact on $\chi$ by the pairing term is not surprising, as it scales as $A^{-1/2}$, much lower than other terms and so scaling must be considered by the user~\cite{Bertolli2012}.  

That highly correlated terms contribute to a better description of nuclear binding should come as no surprise.  The linearity of the correlation coefficient contains no knowledge of non-linear effects in the nuclear interaction, which are necessary for accurate description of nuclei~\cite{Lunney2003,Bethe1965,Machleidt1989,Nazarewicz1994}.  That is, the nuclear energy surface contains non-linear contributions, such as from interactions greater than one-body.  Therefore, simply looking at the linear correlation coefficient will over look the higher-order (i.e. non-linear) changes to the functional's surface that results from the addition of term.  This is seen as well in Fig.~\ref{fig:deltachi}, where a single outlier appears at $\langle |R| \rangle \simeq 0.25$ that does not greatly contribute to reducing $\chi$.  Evaluation of this term indicates a higher-order effect, not identifiable by the correlation test.  So while a term may be highly correlated at first order, the relationship at higher-order behavior is unknown and thus not necessarily redundant.  Such an effect is also seen in Fig.~\ref{fig:deltachi}, where a certain subset of terms with high correlation ($\langle |R| \rangle \simeq 0.9$) provides a reduction in $\chi$ on the order of $\Delta \chi \simeq 0.1$ MeV.

%
%

The converse of missing physical terms is true as well, where the correlation test may retain terms with no physical meaning due to a low correlation coefficient.  One such term which may be added to the formula of Eq.~(\ref{eq:func0}) is 

\begin{equation}
\frac{n_f}{z_f}+\frac{z_f}{n_f}
\label{eq:nonphysical}
\end{equation}

\noindent which is a symmetric ratio of proton and neutron occupations in the highest occupied orbital.  The term in Eq.~(\ref{eq:nonphysical}) has a low average correlation coefficient $\langle |R| \rangle \simeq 0.1$ and so represents a linearly independent direction in the function space.  However, this term has no physical meaning in the context of nuclear binding.  Thus, despite any potential ability to reduce $\chi$ this term should not be considered for use in a formula for nuclear masses.  So while the linearity of  the correlation test does not eliminate its value as an aid in function development, the user must apply context and insight in its use.  The necessary insights may be guided by the application of similar statistical techniques to the nuclear observables produced by energy functionals (see for instance Ref.~\cite{Reinhard2010}).  Such investigations of the key nuclear observables provide a further indication of the physical role of necessary terms that may be rejected due to high correlation.

\section{Conclusions}
\label{conclusions}

Nuclear DFT has proven to be a useful tool for globally describing the ground-state properties of nuclei, yet the construction of the functional
itself poses a significant challenge.  The correlation test introduced in Ref.~\cite{Bertolli2012} provides a computationally inexpensive method of discriminating amongst potential terms to include in  a nuclear mass formula or functional.  In this paper we have provided the necessary mathematical foundation of the method (Sect.~\ref{proof2}) that has hitherto been missing, and quantitative characterization of its performance.

In Sect.~\ref{results} we have developed the benefits of the correlation test by benchmarking the considerably low computational expense.  In Sect.~\ref{results2} we further demonstrated the performance of the correlation test in selecting appropriate terms for the development of a nuclear mass functional.  The inclusion of terms with low correlations led to a reduction in the least-squares error on the order of $\Delta \chi \sim 0.1$ MeV, compared to reductions on the order of $\Delta \chi \sim 0.001$ MeV of highly correlated terms. 

In order to provide a complete characterization of correlation testing, in Sect.~\ref{limitations} we turn to the limitations of the method.  A quantitative description is presented in Sect.~\ref{doi} demonstrating the methods issues in providing a measure of the degree of independence between any two terms in a function.  In Sect.~\ref{nophysics} we further demonstrate the need for physical insight in the use of the correlation test.  The correlation test has no knowledge of the necessary physics, and so the cautious user is not free of the detailed considerations necessary in the field of mass modeling.  Resolving this apparent gap between the use of the correlation test and the necessary physical intuition may be accomplished through application of complimentary work with nuclear observables and other numerical investigations of energy functionals~\cite{Reinhard2010,Kortelainen2012,Gao2013}

Correlation testing is unable to provide direct physical insight on its own, however the computationally inexpensive and systematic method which it provides in selecting functional terms is not only a worthwhile, but essential tool for further work in nuclear mass model development.  With this more rigorous foundation, the method of correlation testing may be elevated from a ``quick-and-dirty" method to simply a ``quick'' method for the systematic development of functional and formulae in the study of nuclear masses.  While the method is computationally cheap and provides a reliable indication of principal axes in a functional space, it is not without limitations.   Improvements on the method may come from a more rigorous treatment of the statistical foundation and an investigation of the effects of adding more than a single term at a time.  The inclusion of rank correlation~\cite{Diaconis1988,Spearman1904,Kendall1938} may be implemented in order to remove the inherent linearity of the test, and is another avenue by which the method can be enhanced.  Furthermore, while our discussion has pertained to the development of nuclear mass models, the method presented here should be of interest to any theoretical investigation that requires the optimization of parameterized models.  


\section*{Acknowledgements}

The author would like to acknowledge the discussions with B. Willett that prompted this work, as well as the critiques and reviews of this paper by  S.V. Paulauskas, K.A. Chipps, T. Kawano and J. Therrien.  The author also thanks anonymous referees for thoughtful comments on the manuscript.  This work was carried out under the auspices of the National Nuclear Security Administration of the U.S. Department of Energy at Los Alamos National Laboratory under Contract No. DE-AC52-06NA25396.

\begin{appendices}
\numberwithin{equation}{section}
\section{Demonstration by Contradiction}

The mathematical foundation for the correlation test as an indication of linear independence is set forth here.  We consider two variables $x_1,x_2$ and two functions $y_1(x_1,x_2)$, $y_2(x_1,x_2)$ which depend on $x_1,x_2$.   We will demonstrate by contradiction that $y_1$ and $y_2$ are linearly independent in $x_1$ and $x_2$ if their covariance is zero.  This differs from statistical independence as $y_1$ and $y_2$ are clearly dependent on the same random variables $x_1$ and $x_2$.

We begin by setting out to show that if: 

\begin{equation}
\cov (y_1,y_2) = 0
\end{equation}

\noindent then $y_1$ and $y_2$ are not linearly dependent.  To demonstrate by contradiction we start by assuming $y_1$ and 
$y_2$ are linearly dependent and attempt to show their covariance can still be zero only with a mathematical contradiction.

Let $y_1,y_2$ be two linear functions of $x_1,x_2$ that are linearly dependent:

\begin{eqnarray}
y_1 &=& ax_1+bx_2 \\
y_2 &=& cx_1+dx_2
\end{eqnarray}

The coefficients $a,b,c,d$ must satisfy the condition for linear dependence:

\begin{equation}
\left| \begin{array}{cc}
a & b \\
c & d  \end{array} \right| = 0
\label{eq:linind}
\end{equation}

\noindent where $|~|$ indicates the determinant.  As we are interested in the application of this method to the development of nuclear functionals and formulae, and for those in which the correlation coefficient is defined, we set the following assumptions:

\begin{enumerate}
\item $\var (x_1),\var (x_2) \ne 0$
\item $y_1,y_2 \in \mathbb{R}$
\item $x_1,x_2 \in \mathbb{R}$
\item $a,b,c,d \in \mathbb{R}$
\end{enumerate}

\noindent We now calculate $\cov (y_1,y_2)$: 

\begin{equation}
\cov (y_1,y_2) = \langle y_1 y_2 \rangle-\langle y_1\rangle \langle y_2 \rangle
\label{eq:covy1y2}
\end{equation}

\noindent where

\begin{eqnarray}
\langle y_1\rangle &=& a\langle x_1\rangle+b\langle x_2\rangle \label{eq:vary1}\\
\langle y_2\rangle &=& c\langle x_1\rangle+d\langle x_2\rangle \label{eq:vary2}\\
\langle y_1 y_2 \rangle &=& ac\langle x_1^2\rangle+bd\langle x_2^2\rangle+(ad+bc)\langle x_1 x_2 \rangle \label{eq:vary1y2}
\end{eqnarray}

Putting Eqs.~(\ref{eq:vary1})-(\ref{eq:vary1y2}) into Eqn~(\ref{eq:covy1y2}) we obtain:

\begin{equation}
\begin{split}
\cov (y_1,y_2) &= ac\langle x_1^2\rangle+bd\langle x_2^2\rangle+(ad+bc)\langle x_1 x_2 \rangle\\
&-ac\langle x_1\rangle^2-bd\langle x_2\rangle^2-(ad+bc)\langle x_1\rangle \langle x_2\rangle
\end{split}
\end{equation}

\noindent Some simplification gives:

\begin{eqnarray}
\cov (y_1,y_2) &=& ac (\langle x_1^2\rangle-\langle x_1\rangle^2)+bd (\langle x_2^2\rangle-\langle x_2\rangle^2) \nonumber \\
&&+(ad+bc)(\langle x_1 x_2\rangle-\langle x_1\rangle \langle x_2\rangle) \nonumber \\
&=& ac \var(x_1)+bd \var(x_2) \nonumber \\
&&+(ad+bc)\cov (x_1,x_2)=0 \label{eq:covgeneral}
\end{eqnarray}

\noindent Utilizing the Eq.~(\ref{eq:linind}), we have the following relationship:

\begin{equation}
ad-bc = 0.
\end{equation}

\noindent Using this relation in Eq.~(\ref{eq:covgeneral}) we may rewrite as

\begin{equation}
\frac{b}{d} c^2 \var (x_1)+bd \var(x_2)+2bc \cov (x_1,x_2) = 0.
\label{eq:subcov}
\end{equation}

We treat Eq.~(\ref{eq:subcov}) as a quadratic equation in $c$, and attempt to solve for the coefficient.

\begin{equation}
\begin{split}
c =& \frac{-b\cov (x_1,x_2)}{\frac{b}{d}\var (x_1)} \\
& \pm \frac{\sqrt{b^2[\cov (x_1,x_2)]^2-b^2\var (x_1)\var(x_2)}}{\frac{b}{d}\var (x_1)}
\end{split}
\label{eq:quad}
\end{equation}

\noindent In order to satisfy the initial assumption $c \in \mathbb{R}$ we must verify the discriminant of Eq.~(\ref{eq:quad}) is non-negative, 

\begin{equation}
b^2[\cov (x_1,x_2)]^2-b^2\var (x_1)\var(x_2) \geq 0
\label{eq:dis}
\end{equation}

\noindent Eq.~(\ref{eq:dis}) simplifies to

\begin{equation}
[\cov (x_1,x_2)]^2 \geq \var (x_1)\var(x_2)
\end{equation}

\noindent However, only in the case of $[\cov (x_1,x_2)]^2 = \var (x_1)\var(x_2)$ can this satisfy the Cauchy-Schwarz inequality~\cite{Steele2004}

\begin{equation}
[\cov (x_1,x_2)]^2 \leq \var (x_1)\var(x_2)
\label{eq:contra1}
\end{equation}

\noindent where the equality only holds true when $x_2$ is a multiple of $x_1$, $x_2 = \beta x_1$.  In the case of an equality in 
Eq.~(\ref{eq:contra1}), we can rewrite $y_1,y_2$:

\begin{eqnarray}
y_1 &=& ax_1+b\beta x_1 = f x_1\\
y_2 &=& cx_1+d \beta x_1 = g x_1
\end{eqnarray}

\noindent where $f=a+b\beta$ and $g=c+d\beta$.  Making use of the properties $\cov (\alpha x,y) = \alpha \cov (x,y)$ and $\cov (x,x) = \var (x)$, the covariance is

\begin{eqnarray}
\cov (y_1,y_2) &=& \cov (fx_1,gx_1)\\
&=&fg\var (x_1)
\label{eq:covx1x1}
\end{eqnarray}

\noindent Where we have assumed $\var(x_1),\var(x_2) \ne 0$, Eq.~(\ref{eq:covx1x1}) is zero only for $fg=0$.  Therefore,

\begin{equation}
\cov (y_1,y_2) = 0
\end{equation}

\noindent only in the case of  $y_1=0$ or $y_2=0$, or $c \notin \mathbb{R}$.  We find a contradiction in our initial assumptions and if $\cov (y_1,y_2) = 0$ then $y_1,y_2$ are not linearly dependent when $a,b,c,d \in \mathbb{R}$, despite a statistical dependence.  The extension of this demonstration to more variables is straightforward and so we omit it here.
\end{appendices}

\end{document}